\documentclass[aps,prb,twocolumn,groupedaddress,superscriptaddress,showpacs]{revtex4-1}
\usepackage{graphicx}
\usepackage{amsmath}
\usepackage{hyperref}
\usepackage{textcomp}
\hypersetup{colorlinks=true, citecolor=blue, urlcolor=blue, linkcolor=blue}

\newcommand {\SPS}    {Sn$_2$P$_2$S$_6$ }
\newcommand {\SPSe}    {Sn$_2$P$_2$Se$_6$ }
\newcommand {\PSPSS}    {(Pb$_y$Sn$_{1-y}$)$_2$P$_2$(Se$_x$S$_{1-x}$)$_6$ }
\newcommand {\PSPS}    {(Pb$_y$Sn$_{1-y}$)$_2$P$_2$S$_6$ }
\newcommand {\PSPSe}    {(Pb$_y$Sn$_{1-y}$)$_2$P$_2$Se$_6$ }
\newcommand {\SPSS}    {Sn$_2$P$_2$(Se$_x$S$_{1-x}$)$_6$ }

\begin{document}

\title{Phase diagram of ferroelectrics with tricritical and Lifshitz points at coupling between polar and antipolar fluctuations}

\author{V.~Liubachko}
\affiliation{Institute for Solid State Physics and Chemistry, Uzhgorod University, Pidgirna Str. 46, Uzhgorod, 88000, Ukraine}
\author{A.~Oleaga}
\affiliation{Departamento de Fisica Aplicada I, Escuela de Ingenieria de Bilbao, Universidad del Pais Vasco UPV/EHU, Plaza Torres Quevedo 1, 48013, Bilbao, Spain}
\author{A.~Salazar}
\affiliation{Departamento de Fisica Aplicada I, Escuela de Ingenieria de Bilbao, Universidad del Pais Vasco UPV/EHU, Plaza Torres Quevedo 1, 48013, Bilbao, Spain}
\author{R.~Yevych}
\affiliation{Institute for Solid State Physics and Chemistry, Uzhgorod University, Pidgirna Str. 46, Uzhgorod, 88000, Ukraine}
\author{A.~Kohutych}
\affiliation{Institute for Solid State Physics and Chemistry, Uzhgorod University, Pidgirna Str. 46, Uzhgorod, 88000, Ukraine}
\author{Yu.~Vysochanskii}
\affiliation{Institute for Solid State Physics and Chemistry, Uzhgorod University, Pidgirna Str. 46, Uzhgorod, 88000, Ukraine}

\date{\today}

\begin{abstract}
Static and dynamic critical behavior of \SPS type ferroelectrics and \PSPSS mixed crystals with line of tricritical points and line of Lifshitz points on the $T-x-y$ phase diagram, which meet at the tricritical Lifshitz point, can be described in a combined BEG-ANNNI model. Such the model considers first and second neighbor interactions for pseudospins in a local three-well potential. Below the temperature of tricritical Lifshitz point, the ``chaotic''' state accompanied by the coexistence of ferroelectric, metastable paraelectric and modulated phases is expected. In addition to the frustration of polar fluctuations near the Brillouin zone center, in \SPS crystal the antipolar fluctuations also strongly develop in the paraelectric phase at cooling to continuous phase transition temperature $T_0$. Here, critical behavior can be described as a crossover between Ising and XY universality classes, what is expected near bicritical points with coupled polar and antipolar order parameters and competing instabilities in $q$-space.  

\end{abstract}

\maketitle

\section{Introduction}
Competing order parameters produce rich phase diagrams of crystals with dipolar ordering in terms of temperature vs external fields, pressure and doping.\cite{b2019_a} Besides, existence of competing ferroelectric, incommensurately modulated and antiferroelectric states with different symmetries in the ground state of the system can be observed. The usual approach to describe the phase diagrams is the Landau-Ginzburg expansion of the free energy density in terms of the order parameters.\cite{b2019_b} At microscopic level, the phase diagrams can be obtained by using mean-field approximations of simplified model Hamiltonians.\cite{b2019_c} At any consideration the main qualitative properties of the phase diagram are determined by investigated system symmetry. 

The active role of several types of mode-mode couplings also is reflected in the complex shape of the local potential for spontaneous polarization fluctuations. It determines the appearance of phase diagrams with different multicritical points that have been intensively studied in the case of oxide materials.\cite{b2019_8,b2019_9,b2019_9_2,b2019_10} The chalcogenide ferroelectrics of \SPS family also present a rich set of scenarios for critical behavior realization\cite{b2019_11,b2019_12} and comparison with the theoretical investigations.\cite{b2019_13,b2019_14}

Ferroelectrics of the \SPS family are promising candidates for application in red and near-infrared spectral diapason photorefraction\cite{b2019_1} and photovoltaics,\cite{b2019_2} for energy storage\cite{b2019_3} and low temperature thermometry,\cite{b2019_d,b2019_d_2} as well as development of multilevel-cell-type memory technology.\cite{b2019_4} Possibilities of ultrafast spontaneous polarization switching\cite{b2019_5} and peculiarities in the ferroelectric ordering and domain structure morphology at nanoscale\cite{b2019_6,b2019_7} are also important for new technologies based on ferroics with linear and nonlinear coupling between unstable lattice polar modes and antipolar or antiferrodistorsive degrees of freedom. 

In \SPS crystals the second order phase transition at $T_0\approx337$~K with lattice symmetry lowering from $P2_1/c$ to $Pc$ has mixed displacive-order/disorder character\cite{b2019_7_1,b2019_7_2} which is manifested by high values of both the transition entropy\cite{b2019_15} $\Delta S=8.6$~JK$^{-1}$mol$^{-1}$ and the dielectric anomaly of the Curie-Weiss constant\cite{b2019_16} $C\approx 10^5$~K. In \SPS crystals under compression, the second order transitions line $T_0(p)$ is monotonously decreases with increasing of temperature and after reaching the tricritical point (TCP) at $p_{\mathrm{TCP}} \approx0.6$~GPa and $T_{\mathrm{TCP}}\approx220$~K, it becomes first order.\cite{b2019_17,b2019_18} At Sn by Pb replacement in \PSPS solid solutions the continuous phase transitions line $T_0(y)$ becomes first order at $y\geq  y_{\mathrm{TCP}}\approx 0.2$ and $T\leq T_{\mathrm{TCP}}\approx 220$~K, and the paraelectric phase becomes stable at cooling down to 0~K for $y\geq 0.6$.\cite{b2019_19,b2019_20} At sulfur by selenium substitution in \SPSS mixed crystals, the incommensurate (IC) phase appears for $x\geq x_{\mathrm{LP}}\approx 0.28$ (where $x_{\mathrm{LP}}$ is the concentration at which the Lifshitz point (LP) appears) on the temperature-concentration diagram.\cite{b2019_12} The virtual paraelectric-ferroelectric transition (inside the IC phase) becomes first order at $x\geq x_{\mathrm{VTCP}}\approx 0.6$ and $T\leq T_{\mathrm{VTCP}}\approx 220$~K.\cite{b2019_11}

It is seen that under any influence, that lowers the ferroelectric phase transition temperature below 220~K, transition character evolves to first order, what has been explained within a model with three-well local potential for the spontaneous polarization fluctuations.\cite{b2019_21} Such shape of the potential energy landscape in \SPS ferroelectrics can be described on the basis of a second order Jahn-Teller effect with nonlinear coupling of polar ($B_u$-symmetry) and fully symmetrical ($A_g$-symmetry) long wave fluctuations. This local potential was found by first principles calculations\cite{b2019_22} and is related to $5s^2$ electrons lone pair stereoactivity of Sn$^{2+}$ cations.\cite{b2019_22,b2019_23} The charge disproportionation of phosphorous cations $P^{3+} + P^{5+} \leftrightarrow P^{4+} + P^{4+}$ also originates an electronic contribution to the spontaneous polarization.\cite{b2019_21} The evolution of pseudospin distribution in three-well local potential while decreasing the temperature was found by MC simulations,\cite{b2019_22} and it is comparable with experimental data of $^{31}P$ NMR spectroscopy.\cite{b2019_24}

In addition to the microscopic atomistic consideration, the nature of the phase transition in \SPS crystals was also analyzed within the framework of the thermodynamic three-state Blume-Emery-Griffth (BEG) model\cite{b2019_25,b2019_26} that supposes 0, +1 and -1 values for local pseudospins and predicts the TCP presence on a temperature-pressure or temperature-composition phase diagram of mixed crystals. The TCP presence and the lowering of the phase transition temperature till 0~K are determined by pressure or composition evolution of the local three-well potential shape (change of the energy difference between central and side wells) and was analyzed in the approach of the quantum anharmonic oscillators (QAO) model for \SPS crystals.\cite{b2019_21} The calculated spectra of pseudospin fluctuations in the framework of the QAO model correlate well with the observed Raman spectra transformation in the ferroelectric phase with appearance of low-energy modes.\cite{b2019_27} Besides, the \SPS crystal lattice anharmonicity on example of thermal expansion temperature anomaly was successfully described using the QAO model.\cite{b2019_28}

The appearance of the IC phase in \SPS family crystals can be explained using the discrete axial Ising model --- ANNNI model,\cite{b2019_29} that consider short range interaction $J_1>0$ between the nearest neighbors, and interaction $J_2<0$ between next-nearest neighbors. The ratio of these interactions $\lambda= - J_2/J_1$ changes at sulfur by selenium substitution in \SPSS  mixed crystals --- the IC phase appears at the LP with coordinate $\lambda= 0.25$. It was estimated\cite{b2019_16} that  $\lambda$ changes from 0.23 for composition with $x = 0$ to 0.3 for $x = 1$. 

Discrete ANNNI model introduces into consideration two sublattices of \SPS type crystals structure with two formula units in the elementary cell. They were involved in the description of the IC phase, for which the wave vector of the spontaneous polarization transverse space modulation wave lies on the monoclinic symmetry plane near [001] direction.\cite{b2019_16} The spontaneous polarization vector also lies in the monoclinic symmetry plane (010) and oriented near [100] direction. Experimental data of neutron scattering for \SPSe crystal demonstrate flexoelectric coupling between soft optical and acoustic phonon branches near the Brillouin zone center as their repulsion due to the same symmetry along $q_z$ direction of reciprocal space.\cite{b2019_30} This coupling has also been observed in temperature anomalies of ultrasound and hypersound velocity.\cite{b2019_31} In the continuous approximation, the linear flexoelectric interaction between optical and acoustic phonons is related to the Lifshitz like invariant $(dP_x/dz)u_{xz}$ in the thermodynamic potential.\cite{b2019_32,b2019_33} Its growth at crossing from \SPS to \SPSe causes by a higher covalence of chemical bonds in selenide compound.\cite{b2019_34}

The neutron scattering data\cite{b2019_35} also show the presence of a flat lowest-energy transverse optical branch along $q_y$ direction in the paraelectric phase of \SPS crystal. Developed diffuse X-ray scattering along $q_y$ direction in the paraelectric phase near temperature $T_0$ also confirms an important role of the possibility of short-wave antipolar fluctuations.\cite{b2019_36} 

In this paper we use the model approaches for the description of the phase diagram topology of \SPS based ferroelectric mixed crystals as well as of the critical behavior of their thermodynamic and dynamic properties, considering the important role of developed polar fluctuations coupling with fully symmetrical and antipolar fluctuations together with frustrations whenever there is a competition between different instabilities in reciprocal $q$-space.

\section{Phase diagram with tricritical Lifshitz point}

In order to build the $T-\lambda-\Delta$ phase diagram for ferroelectrics in the system Sn(Pb)$_2$P$_2$S(Se)$_6$, we have used the Blume-Capel spin-1 Ising model (which is a simpler version of BEG model) with competing interactions.\cite{b2019_37,b2019_38} Such combined BC-ANNNI model displays a multicritical behavior such as tricriticality as well as the possibility of the presence of a LP. The tricritical Lifshitz point (TCLP) is predicted at the meeting of TCPs and LPs lines.\cite{b2019_39,b2019_39_2,b2019_40} In the used model\cite{b2019_37} (as for original ANNNI model\cite{b2019_29}) the relation between two interactions is written in the form $\lambda= - J_2/J_1$, where $J_1$ is the effective first neighbors’ positive interaction and $J_2$ is the negative coupling of next nearest neighbors. The paraelectric-ferroelectric second order transitions line can be found as\cite{b2019_37}
\begin{equation}
\lambda=1-\frac{t}{1+0.5e^{\Delta/t}}\label{eq1}
\end{equation}
\noindent and from the paraelectric phase into a modulated one continuous transitions line
\begin{equation}
\lambda=\frac{t}{1+0.5e^{\Delta/t}}.\label{eq2}
\end{equation}
\noindent Here $t = T/J_1$, and $\Delta= \delta/J_1$. The parameter $\delta$ is related to the single-ion terms.\cite{b2019_21,b2019_25}
 At $\lambda = 0.25$ the paraelectric borders obtained from (\ref{eq1}) and (\ref{eq2}) meet at a LP. Considering the above-mentioned conditions, the line of LPs can be found by means of the following equation
 \begin{equation}
	\Delta= t_{\mathrm{LP}} \ln{ \frac{ 1-2t_{\mathrm{LP}}}{t_{\mathrm{LP}} } }. \label{eq3}
\end{equation}

\begin{figure}[!htb]
\includegraphics*[width=0.9\columnwidth]{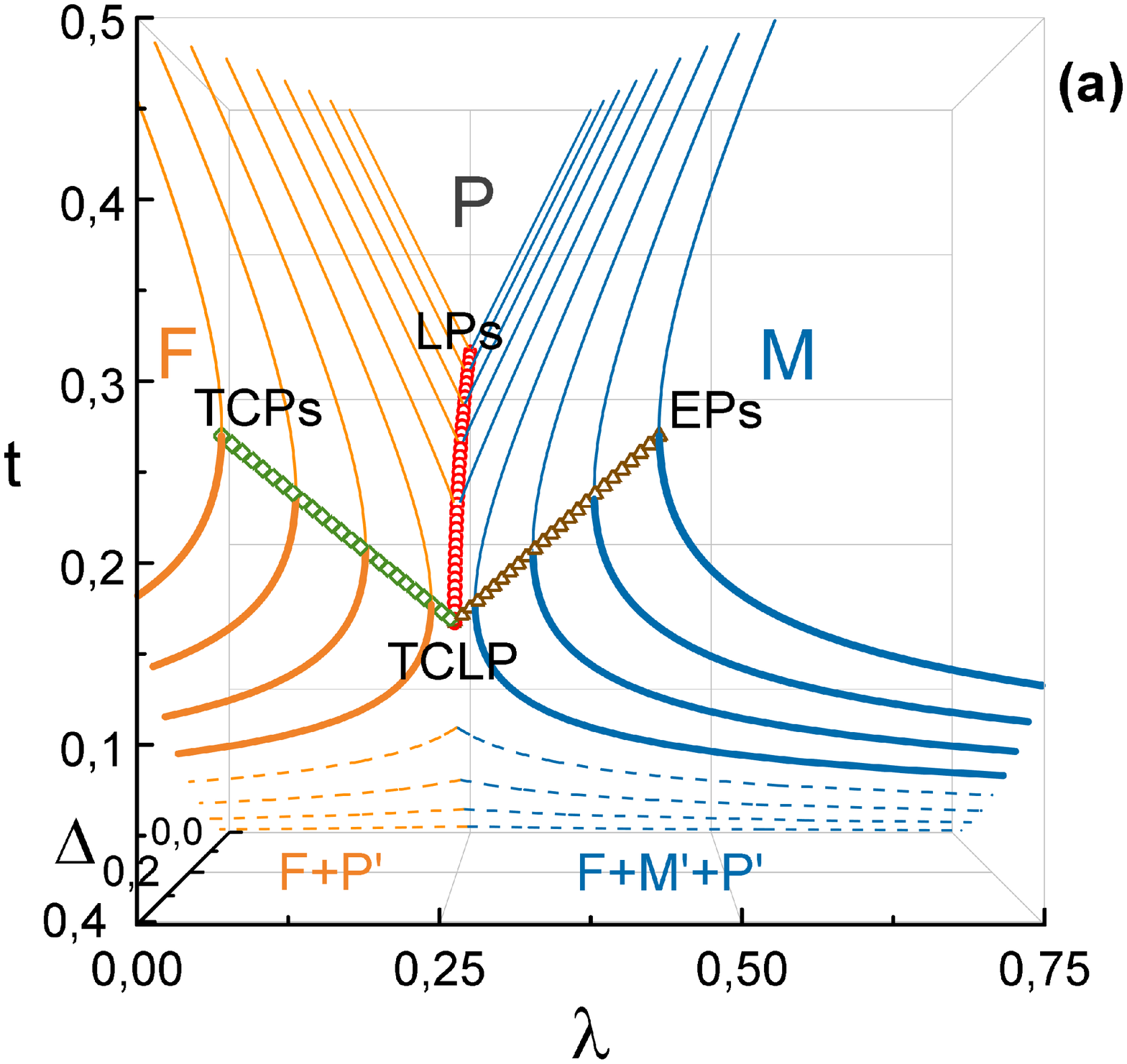}
\includegraphics*[width=0.9\columnwidth]{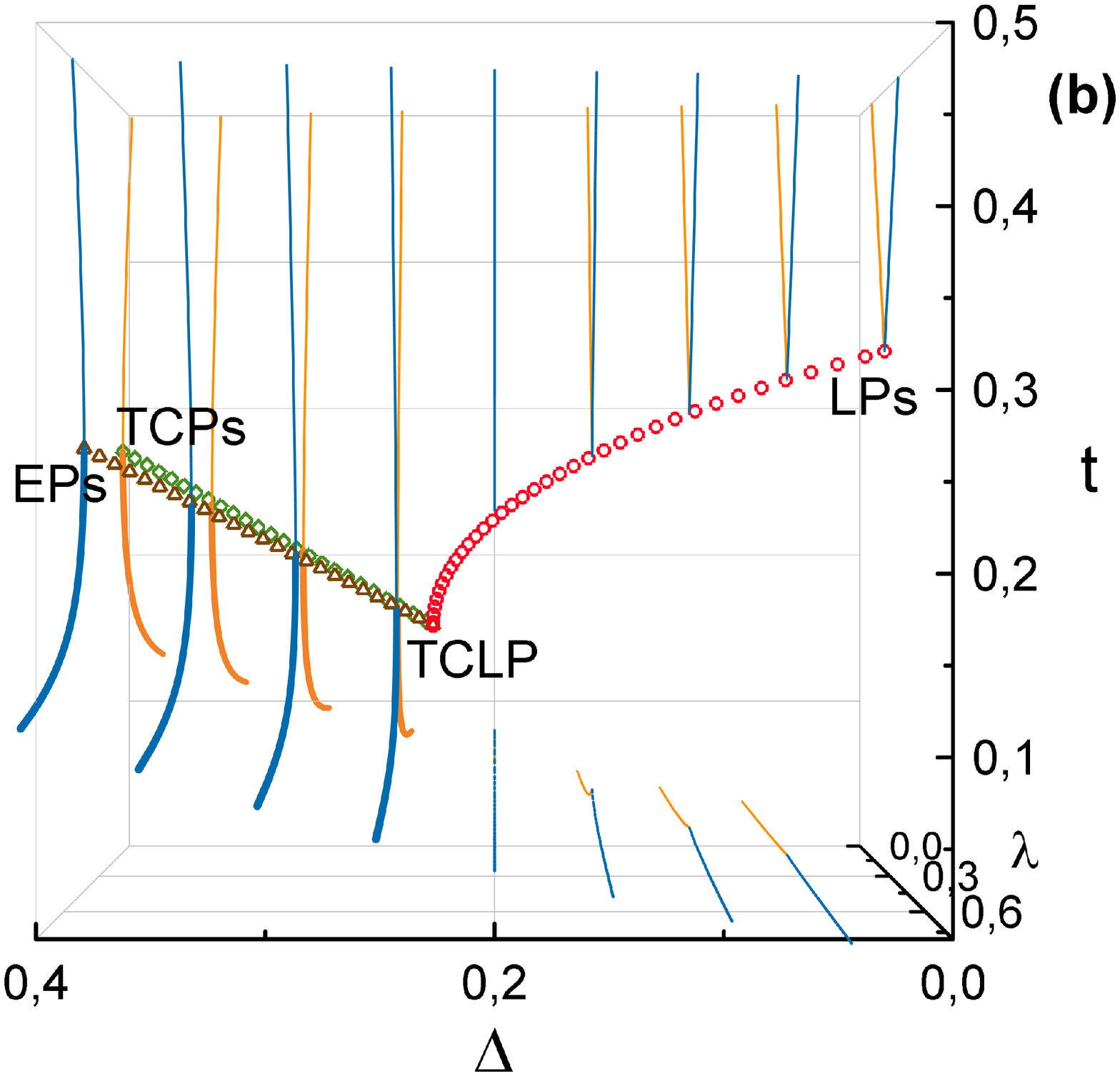}
 \caption{(a) Front and (b) side view on calculated $t-\lambda-\Delta$ diagram. Solid lines show paraelectric-ferroelectric (orange) and paraelectric-modulated (blue) borders, Lifshitz points (LPs) line (red circles), tricritical points (TCPs) line (green squares), end points (EPs)  line (brown triangles). Thin lines indicate the second order phase transitions while thick lines are the first order ones. Letters F, P and M correspond to ferroelectric, paraelectric and modulated phases. The coexistence of ferroelectric and metastable paraelectric (F+P\textquotesingle) and ferroelectric, metastable paraelectric and commensurate modulated phases (F+M\textquotesingle+P\textquotesingle) below the dashed lines are also shown.\label{fig1}}
\end{figure}

After reaching a value $\Delta\approx 0.231$ the paraelectric borders begin to split and there is no LP anymore. Near this $\Delta$ value the LPs line coincides with the TCPs line and the end points line.\cite{b2019_37} The calculated $t-\lambda-\Delta$ phase diagram is depicted in Fig.~\ref{fig1}.

Let’s compare the just obtained theoretical phase diagram with the experimentally determined temperature-concentration phase diagram for \PSPSS ferroelectric mixed crystals.\cite{b2019_16} In order to do it we need to translate it into $t-\lambda-\Delta$ coordinates. The experimentally observed\cite{b2019_41} TCLP for (Pb$_{0.05}$Sn$_{0.95}$)$_2$P$_2$(Se$_{0.28}$S$_{0.72}$)$_6$ at $T_{\mathrm{TCLP}}\approx 259$~K in $t-\lambda-\Delta$ coordinates will correspond to the next position: $t = 0.158249$, $\lambda= 0.5$, $\Delta = 0.23105$. The LP with the composition  Sn$_2$P$_2$(Se$_{0.28}$S$_{0.72}$)$_6$ at temperature $T_{\mathrm{LP}}\approx 284$~K (Ref.~\onlinecite{b2019_12}) will lie on the line of LPs obtained from eq.~(\ref{eq3}) at $t = 0.17345$, $\lambda= 0.5$, $\Delta= 0.22997$. For mixed \SPSS crystals in the framework of the ANNNI model, a linear variation of the $\lambda$ parameter with composition $x$ was assumed with the values 0.23, 0.25 and 0.30 for Sn$_2$P$_2$S$_6$, Sn$_2$P$_2$(Se$_{0.28}$S$_{0.72}$)$_6$ and Sn$_2$P$_2$Se$_6$, respectively.\cite{b2019_16} At the constant value $\Delta= 0.22997$, the coordinates for these concentrations on the $t-\lambda-\Delta$ diagram are the following: $t_0 = 0.20582$, $\lambda= 0.23$ for \SPS with $T_0 = 337$~K; $t_c = 0.13436$, $\lambda_c = 0.30$, and $t_i = 0.11799$, $\lambda_i = 0.30$ for \SPSe with $T_c = 220$~K, and $T_i = 193$~K.\cite{b2019_16}

\begin{figure}[!htb]
\includegraphics*[width=\columnwidth]{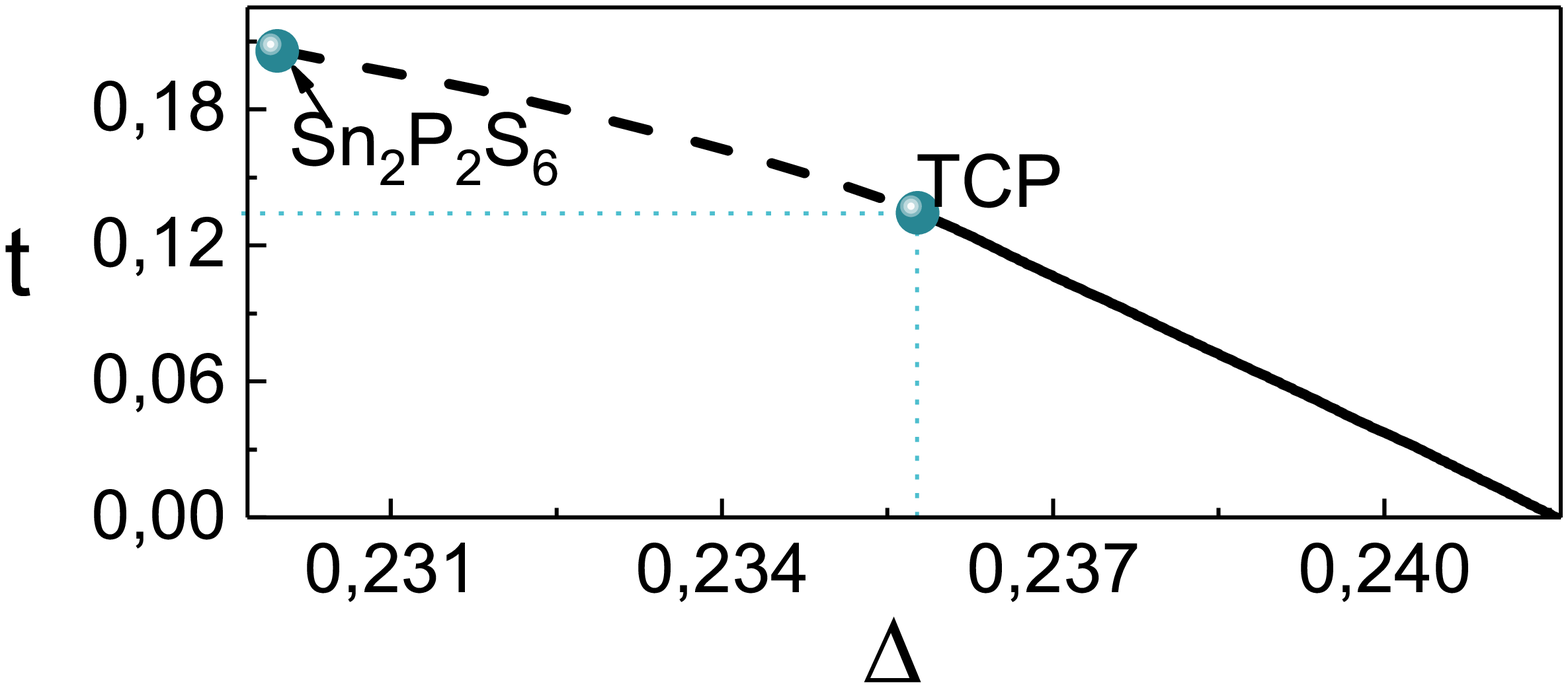}
 \caption{Dependence of phase transition temperature on the single-ion term in dimensionless $t-\Delta$ coordinates calculated in the mean-field approximation on the BEG model.\cite{b2019_21,b2019_42} Dashed and solid lines indicate second and first order transitions, respectively, that meet at the tricritical point.\label{fig2}}
\end{figure}

At tin by lead substitution the shape of local three-well potential changes, and the coordinates of the TCP in the \PSPS mixed crystals in mean-field approximation of BEG model\cite{b2019_42} can be determined at condition of linear variation of $\Delta$ in dependence of $y$, at unchanged intercell interactions. According to the earlier performed analysis for \PSPS mixed crystals,\cite{b2019_21} the calculated $t-\Delta$ diagram (at $\lambda = 0.23$) is shown in Fig.~\ref{fig2}. It was found that the TCP has coordinates $\lambda=0.23$, $t=0.13436$, $\Delta=0.23577$.

\begin{figure}[!htb]
\includegraphics*[width=0.9\columnwidth]{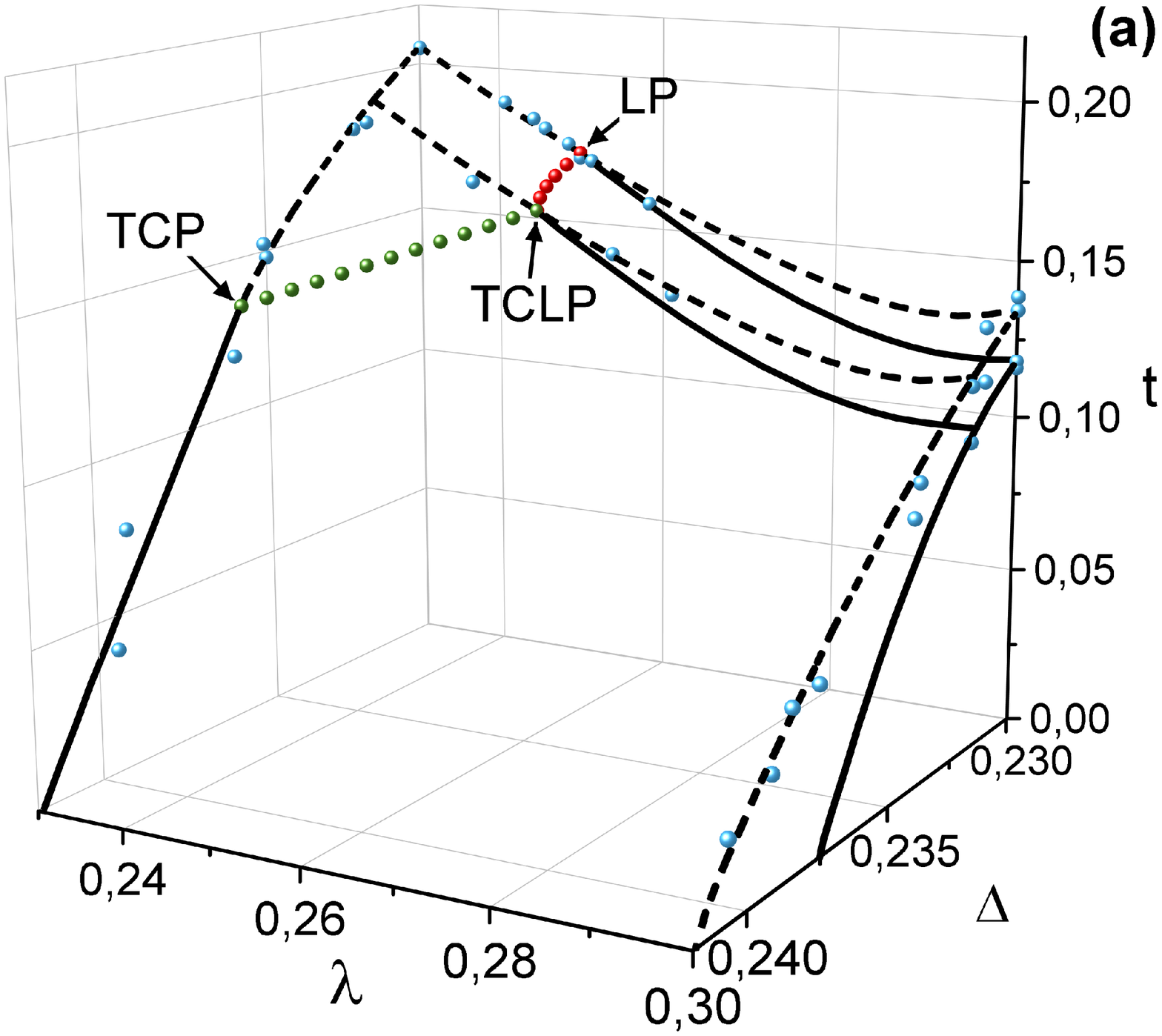}
\includegraphics*[width=0.9\columnwidth]{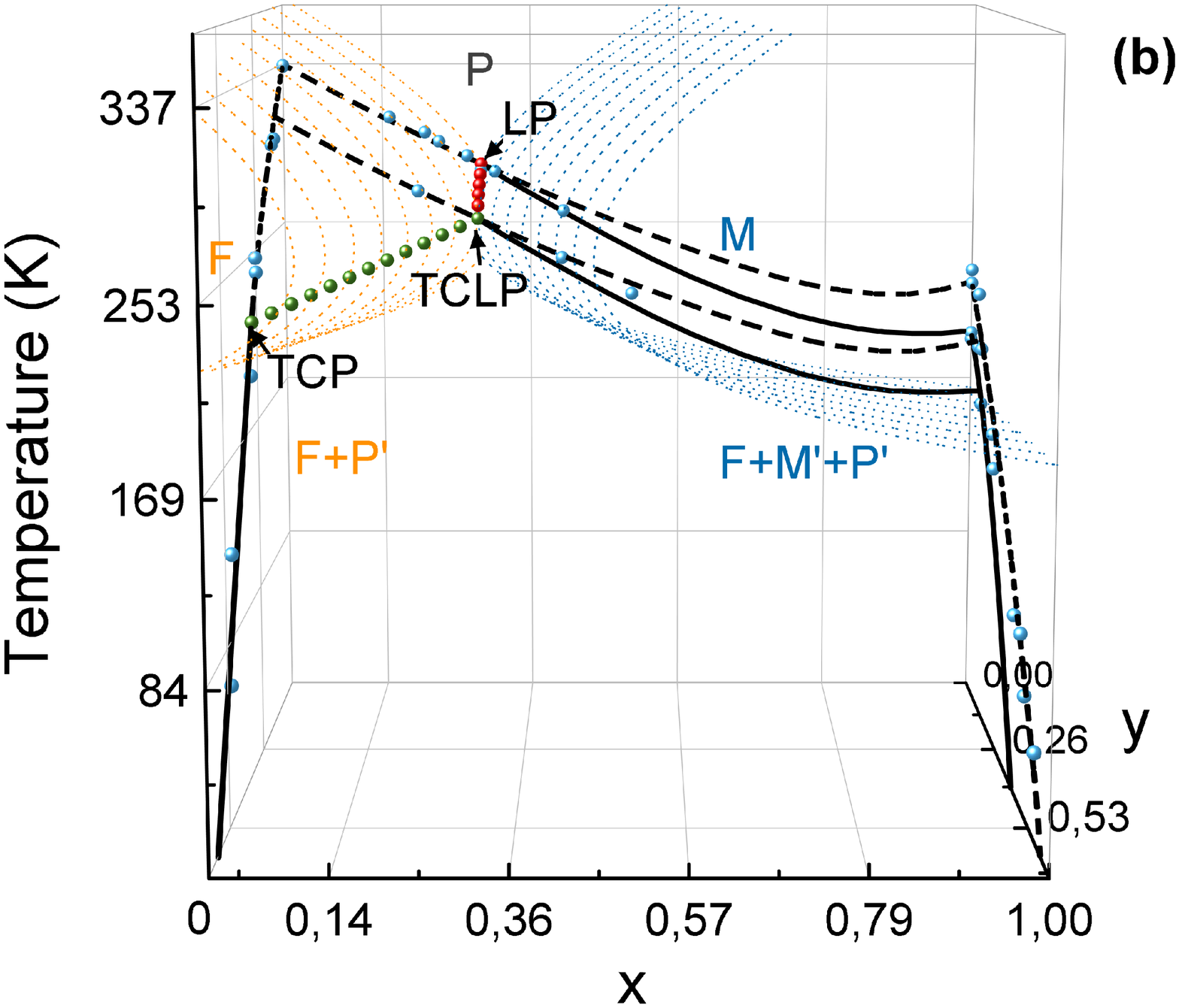}
 \caption{(a) Experimental phase diagram in $t-\lambda-\Delta$ coordinates for \PSPSS ferroelectrics  and (b) front view of both experimental and calculated phase diagrams in temperature-concentration $T-x-y$ coordinates. Dashed and solid lines denote second order and first order phase transitions, respectively. Green and red spheres correspond to tricritical points line and Lifshitz points line, blue spheres to the calorimetric data.\cite{b2019_20,b2019_41,b2019_43,b2019_44,b2019_45,b2019_46,b2019_47}\label{fig3}}
\end{figure}

As the inter-site interaction $J_1$ is an almost unchanged quantity,\cite{b2019_21} we assume that, on increasing the lead concentration in \PSPSe mixed crystals (with $\lambda = 0.3$), the value of $\Delta$ will change in the same way as in \PSPS solid solutions. Accordingly, the experimental phase diagram in $t-\lambda-\Delta$ coordinates is presented in Fig.~\ref{fig3}.

From the calculated $t-\lambda-\Delta$ diagram, it follows that the LPs line terminates at TCLP, and this multicritical point can be considered as a “Lifshitz end point”.  The LPs line in TCLP splits into the tricritical points line and the end points line. At big $\Delta$ parameter, the paraelectric-modulated critical line ends at the end point.\cite{b2019_37}

\begin{figure}
\includegraphics*[width=0.9\columnwidth]{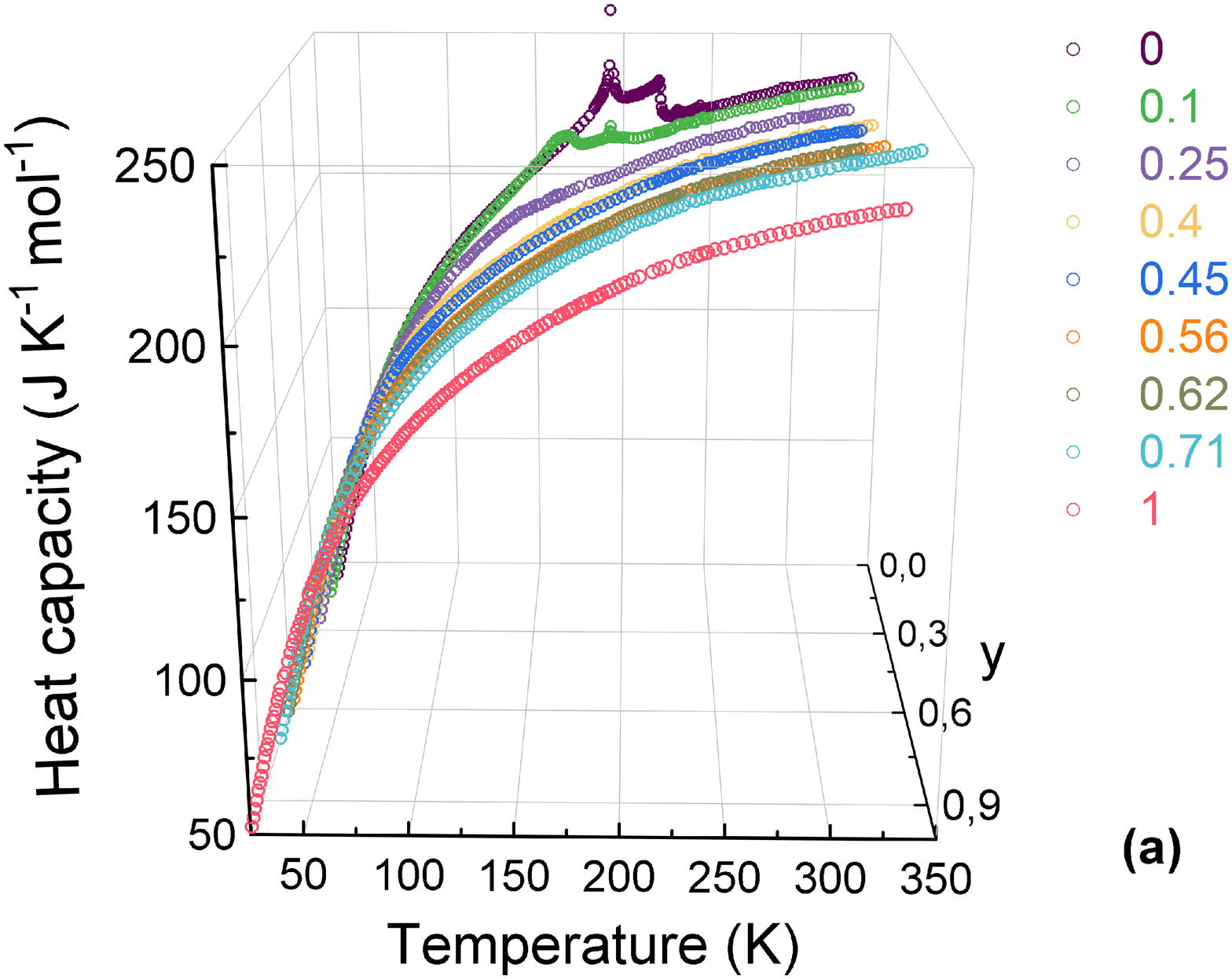}
\includegraphics*[width=0.9\columnwidth]{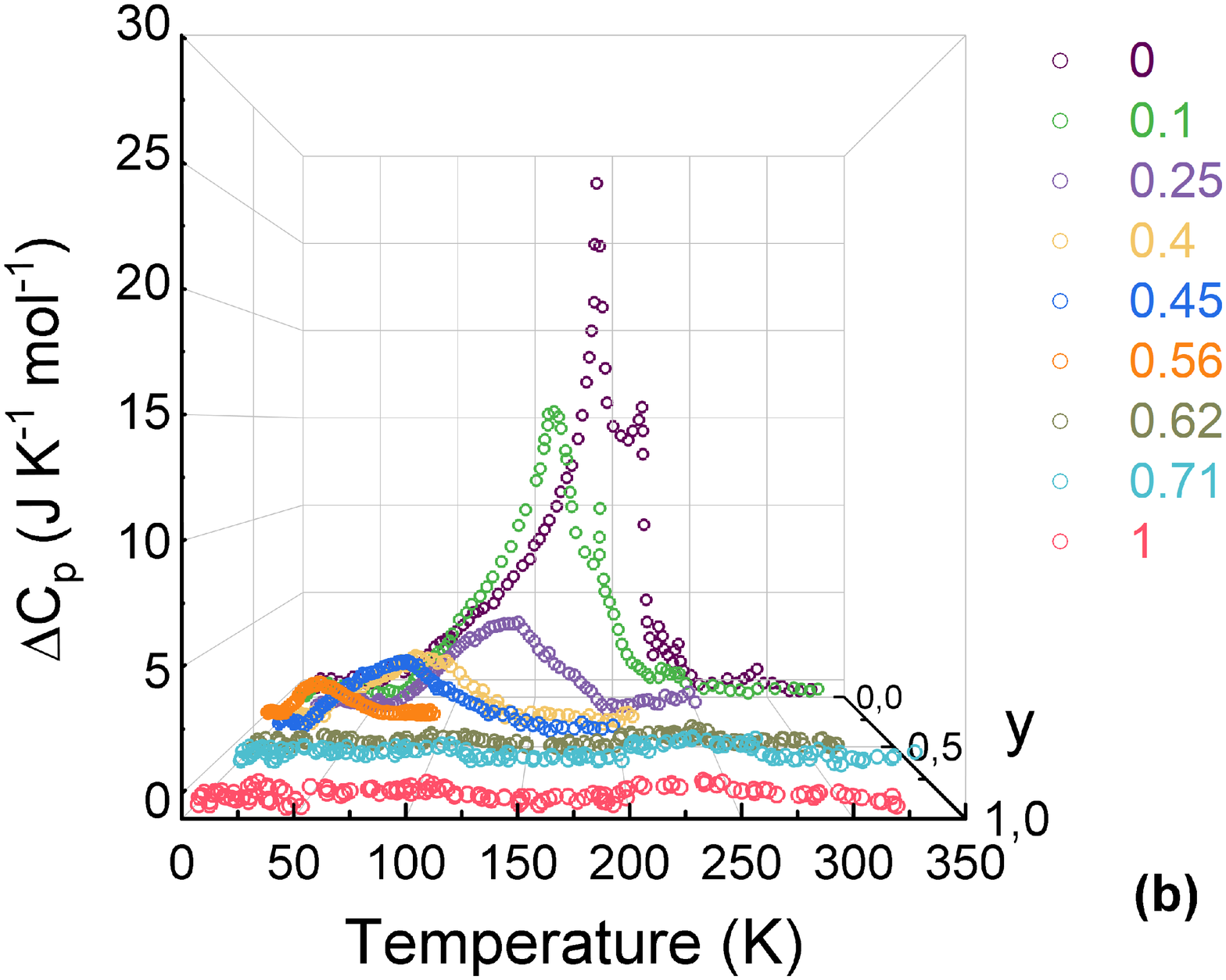}
 \caption{Temperature dependence of (a) heat capacity according to experimental data\cite{b2019_48} and (b) related to the phase transition excess heat capacity for\PSPSe crystals.\label{fig4}}
\end{figure}

\begin{figure*}
\includegraphics*[width=\textwidth]{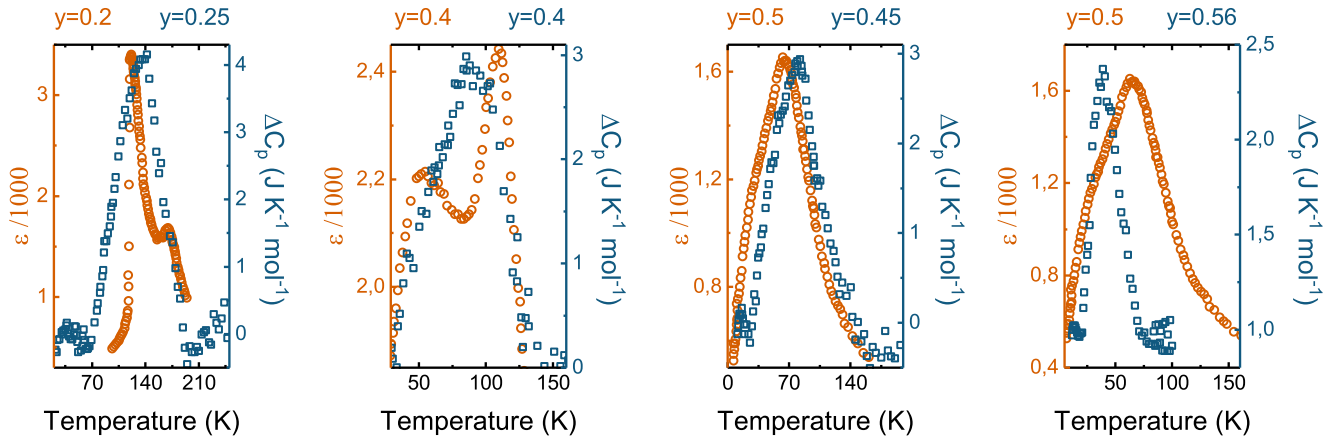}
 \caption{Comparison of the anomalies for low frequency dielectric susceptibility (according to data\cite{b2019_d,b2019_d_2,b2019_49}) and excess heat capacity (according to data\cite{b2019_48}) in the region of the phase transitions for \PSPSe crystals.\label{fig5}}
\end{figure*}

For high lead concentrations ($y > 0.2$), at low temperatures a “chaotic” state can be observed.\cite{b2019_37} This state presents a mixture of paraelectric, ferroelectric and modulated phases. Such peculiarity can be seen on the excess heat capacity $\Delta C_p$ and anomalous temperature dependencies of dielectric susceptibility $\varepsilon'$ in \PSPSe crystals, according to recent investigations.\cite{b2019_d,b2019_d_2,b2019_48,b2019_49} For small lead concentrations, there are clear anomalies of $\Delta C_p(T)$ and  $\varepsilon'(T)$ at paraelectric-incommensurate ($T_i$) and incommensurate-ferroelectric ($T_c$) phase transitions (see Fig.~\ref{fig4}, \ref{fig5}). However, for $y \geq 0.2$ the $\Delta C_p(T)$ and $\varepsilon'(T)$ anomalies in the vicinity of the lock-in transition ($T_c$) are strongly smeared. Such chaotization can be related to a synergy of frustration effects and nonlinearity of the system with the three-well local potential.


Therefore, the ANNNI model with two structural sublattices reflects the main properties of ferroelectrics that are related to the LP and IC phase on their phase diagram. For the description of a presence of a TCP on the phase diagram of \SPS family ferroelectrics, two structural sublattices were also considered in the frame of BEG model.\cite{b2019_25,b2019_26} This three-state model is based on two order parameters --- dipolar and quadrupolar ($B_u$ and $A_g$ symmetry variables in the case of Sn$_2$P$_2$S$_6$).\cite{b2019_22,b2019_23} Therefore, in addition to dipole-dipole inter-site coupling, the quadrupole-quadrupole interaction can also be important. In the case of two sublattices, taking into account the quadrupole-quadrupole interactions can complicate the topology of the temperature-pressure (composition) phase diagram. On such diagram, in addition to (or instead of) the presence of a TCP, another multicritical points can appear, such as a triple point, a critical end point, a bicritical point, and a tetracritical point.\cite{b2019_26} In the case of negative quadrupole-quadrupole coupling at temperature lowering, in addition to the ferroelectric state, ferrielectric and antiquadrupolar phases can also appear.\cite{b2019_26} With growth of positive quadrupole-quadrupole coupling, this TCP transforms into a triple point. In such case, at cooling in the paraelectric phase the first order transition with change of quadrupolar order parameter occurs, and with further temperature lowering the second order transition into ferroelectric state is observed.\cite{b2019_26}

\section{Polar and antipolar fluctuations coupling}

The above mentioned examples of possible complications of experimentally observed phase diagrams for objects with complex local potential for spontaneous polarization fluctuations demonstrate the proximity of the phase transitions to higher order multicritical points, like a tetracritical point, and give evidence of the importance of higher order invariants in the thermodynamic potential of the investigated ferroelectric crystals. Such possibility was earlier demonstrated\cite{b2019_50,b2019_51} when the incommensurate phase properties near lock-in transition into the ferroelectric phase in \SPSe crystals were theoretically explained.
\begin{figure*}[!htb]
\includegraphics*[height=2.in]{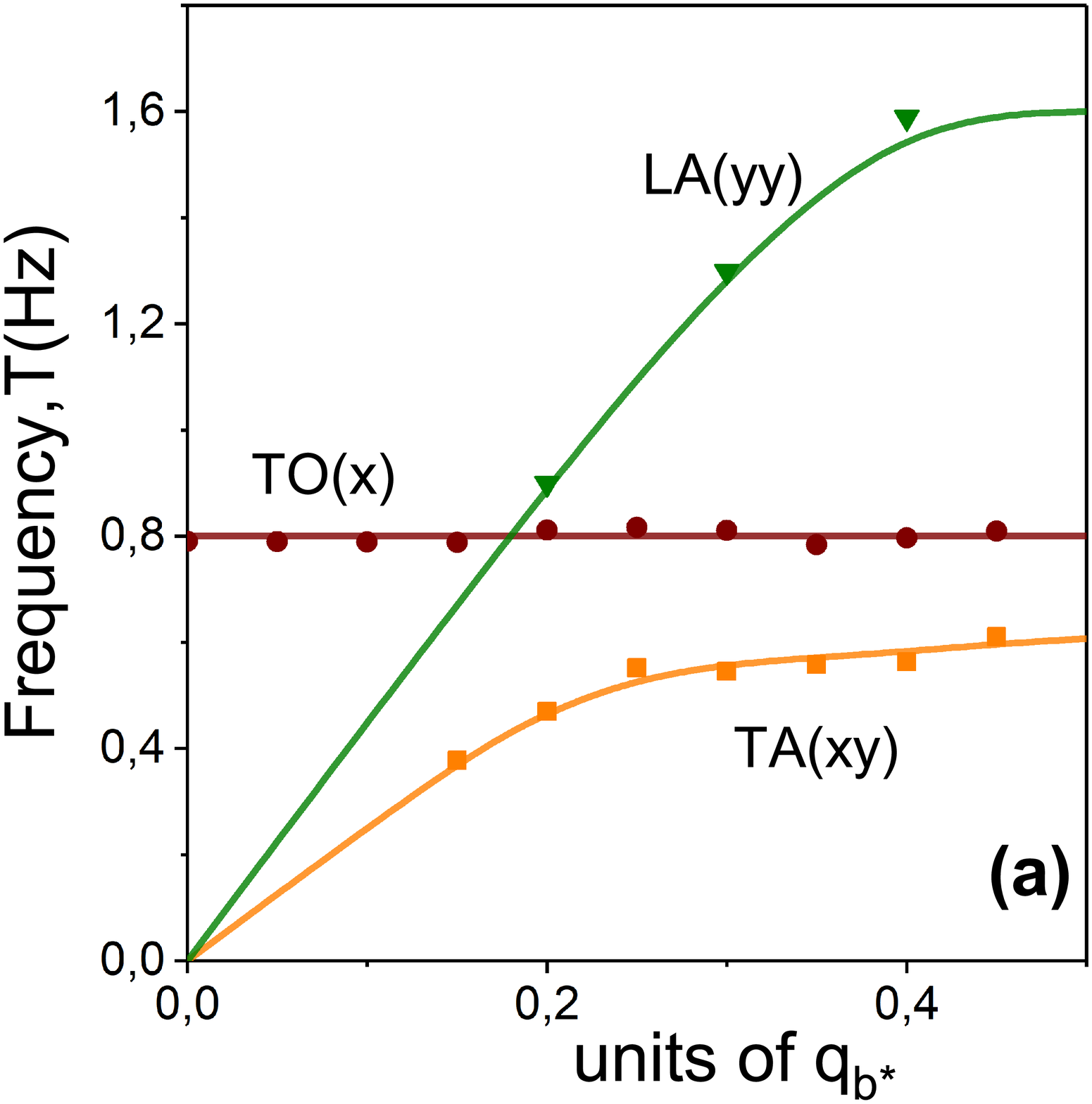}\hfill
\includegraphics*[height=2.in]{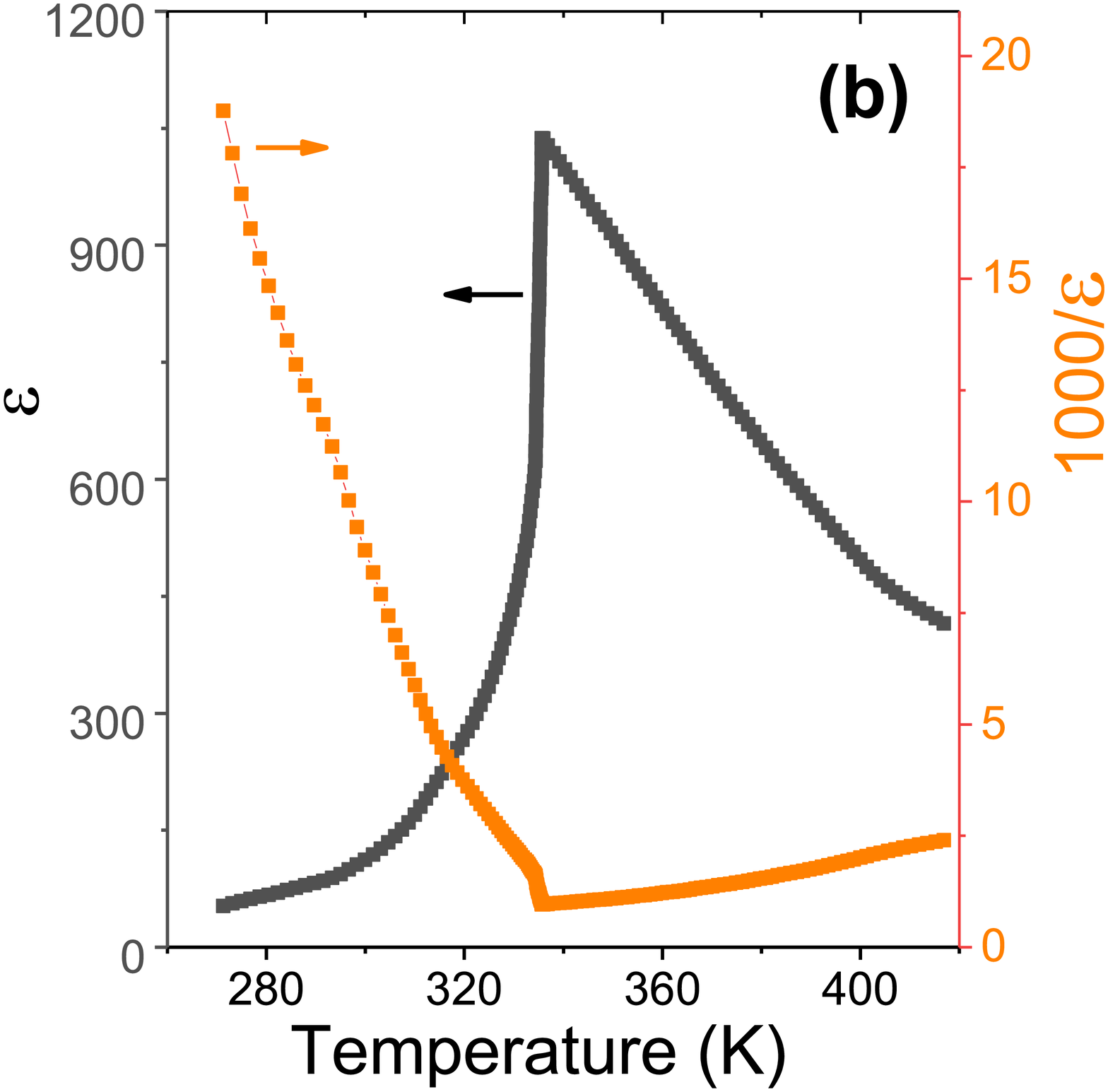}\hfill
\includegraphics*[height=2.in]{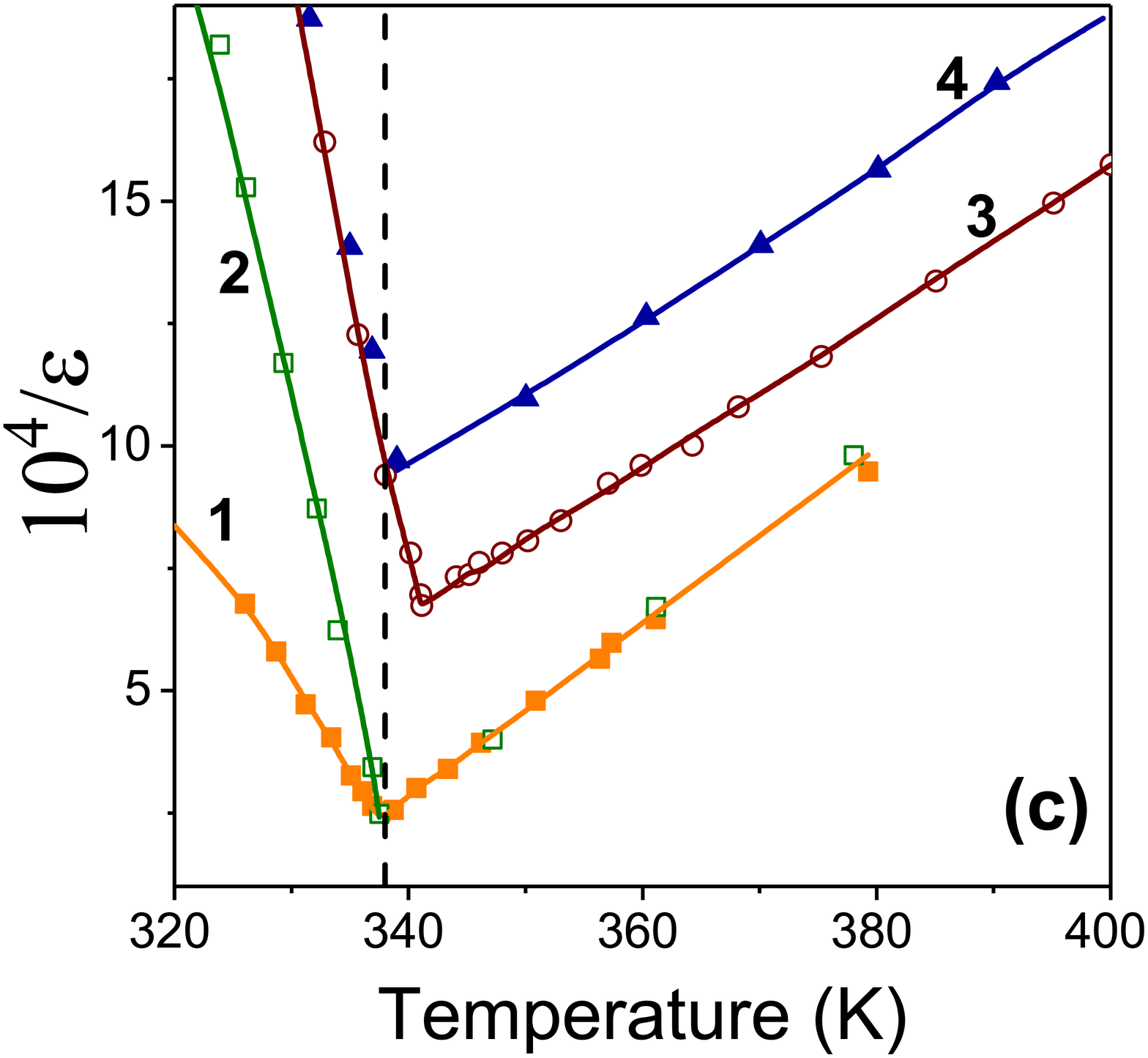}\\
\begin{minipage}[c]{0.3\textwidth}
\includegraphics*[height=2.in]{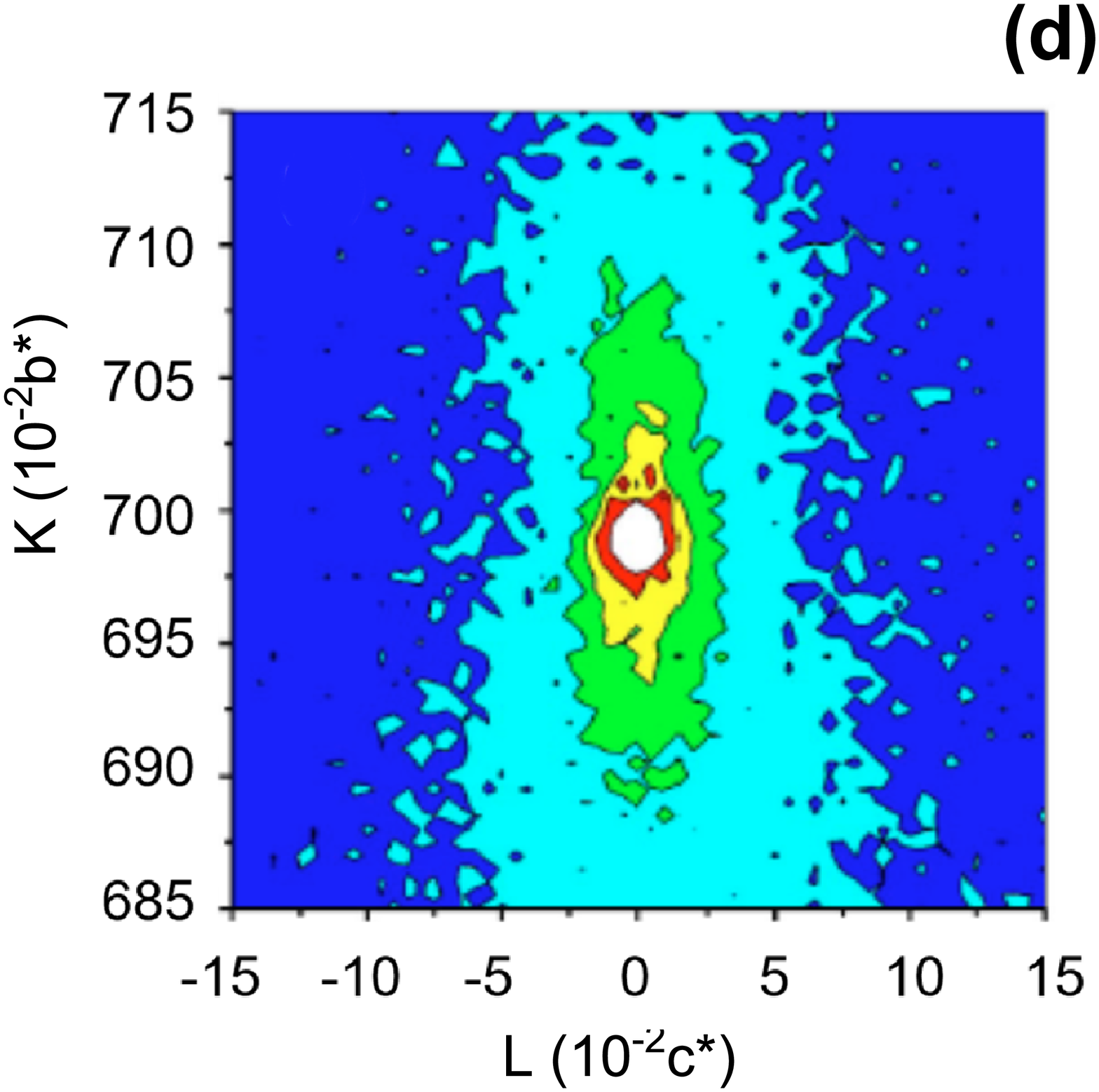}
\end{minipage}\hfill
  \begin{minipage}[c]{0.63\textwidth}
   \caption{(a) Determined by neutron scattering at 440~K transverse soft optical $TO(X)$, acoustic longitudinal $LA(XX)$ and transverse $TA(XY)$ phonon branches along $q_y$ direction of \SPS crystal monoclinic Brillouin zone\cite{b2019_35}; (b) temperature dependence of the dielectric susceptibility and his reciprocal from submillimeter soft optical mode contribution near the phase transition in \SPS crystal\cite{b2019_52}; (c) comparison of the inverse of the dielectric susceptibility temperature dependence in \SPS crystal\cite{b2019_53} for 20~MHz --- 1, 4~GHz --- 2, 27~GHz --- 3 and submillimeter soft optical mode contribution --- 4; (d) diffuse X-ray scattering at $T_0 + 2$~K in the paraelectric phase of \SPS crystal in  the plane (0{\it KL}) of Brillouin zone.\cite{b2019_36} \label{fig6}}
  \end{minipage}
\end{figure*}

For \SPS ferroelectrics, the experimental data of neutron scattering, dielectric susceptibility and hysteresis loops investigations present rich information about the complex character of the phase transitions. The neutron scattering data\cite{b2019_35} show the presence of a flat lowest-energy transverse optical branch along $q_y$ direction at $T=440$~K in the paraelectric phase [see Fig.~\ref{fig6}(a)]. This phonon branch is polarized near [100] direction and softens at cooling. The polar soft optical mode, according to the submillimeter spectroscopy data near $10^{12}$~Hz,\cite{b2019_52} contributes only $\Delta \varepsilon'= 1000$ to the dielectric susceptibility maximum near $T_0 \approx 337$~K [see Fig.~\ref{fig6}(b)]. At frequency lowering till $10^7$~Hz, the dielectric susceptibility near $T_0$ rises to $10^4-10^5$ [see Fig.~\ref{fig6}(c)],\cite{b2019_53} and at lower frequency obeys the Curie-Weiss law $\varepsilon'= C(T-T_0)^{-1}$  with $C \approx 0.6\cdot 10^5$~K. 

On cooling in the paraelectric phase of \SPS crystal, not only long wave polarization fluctuations are developed near the Brillouin zone center, which are proportional to the reciprocal of the polar soft optical mode (with $B_u$ symmetry at $q\rightarrow 0$) frequency square. Near the Brillouin zone edge, at $q_y =\pi /b$, also critical growth occurs for fluctuations of atomic motions that are related to the eigenvector of short length phonons from the soft optical branch $B(q_y)$. The development of the aforementioned fluctuations was observed directly [see Fig.~\ref{fig6}(d)]\cite{b2019_36} by inelastic X-ray scattering at $T_0 + 2$~K.

The high value of the low frequency dielectric susceptibility on cooling to $T_0$ is obviously related to crystal lattice anharmonicities. It was earlier supposed\cite{b2019_22,b2019_23} that the local three-well potential follows from a nonlinear interaction of polar $B_u$ modes with fully symmetrical $A_g$ modes (like $A_gB_u^2 + A_g^2B_u^2$) at the Brillouin zone center. But on the matter of the soft optical mode flatness in $q_y$ direction,\cite{b2019_35} the nonlinear phonon-phonon interaction can be obviously realized with the involvement of phonons from different points of the Brillouin zone. In the simplest approach, such possibility can be incorporated in QAO model\cite{b2019_4,b2019_21,b2019_27,b2019_28} that is based on a three-well on-site potential involving first and second neighbors inter-site interactions. Here we consider nonlocal interactions in $q_y$ direction (instead of above considered frustration of $J_1$ and $J_2$ inter-site interactions that are related to the incommensurate phase appearance with modulation wave vector $q_z$). The phase diagram calculated with such model contains a tetracritical point at which two second order transitions lines (from paraelectric into ferroelectric phase and between paraelectric and antipolar phases) intersect. It was found\cite{b2019_4} that for \SPS crystals below $T_0$ coexistence of antipolar and ferroelectric phases can be presented. 

The assumption that in \SPS crystals the phase transition at $T_0 \approx 337$~K is placed near the tetracritical point agrees with the previous discussion based on the BEG model, for which the phase diagram with a TCP can be complicated by the presence of bicritical or tetracritical points.\cite{b2019_26} Thus, the critical behavior of \SPS crystals near temperature $T_0$ requires special attention. 

\begin{figure*}[!htb]
\includegraphics*[width=0.9\textwidth]{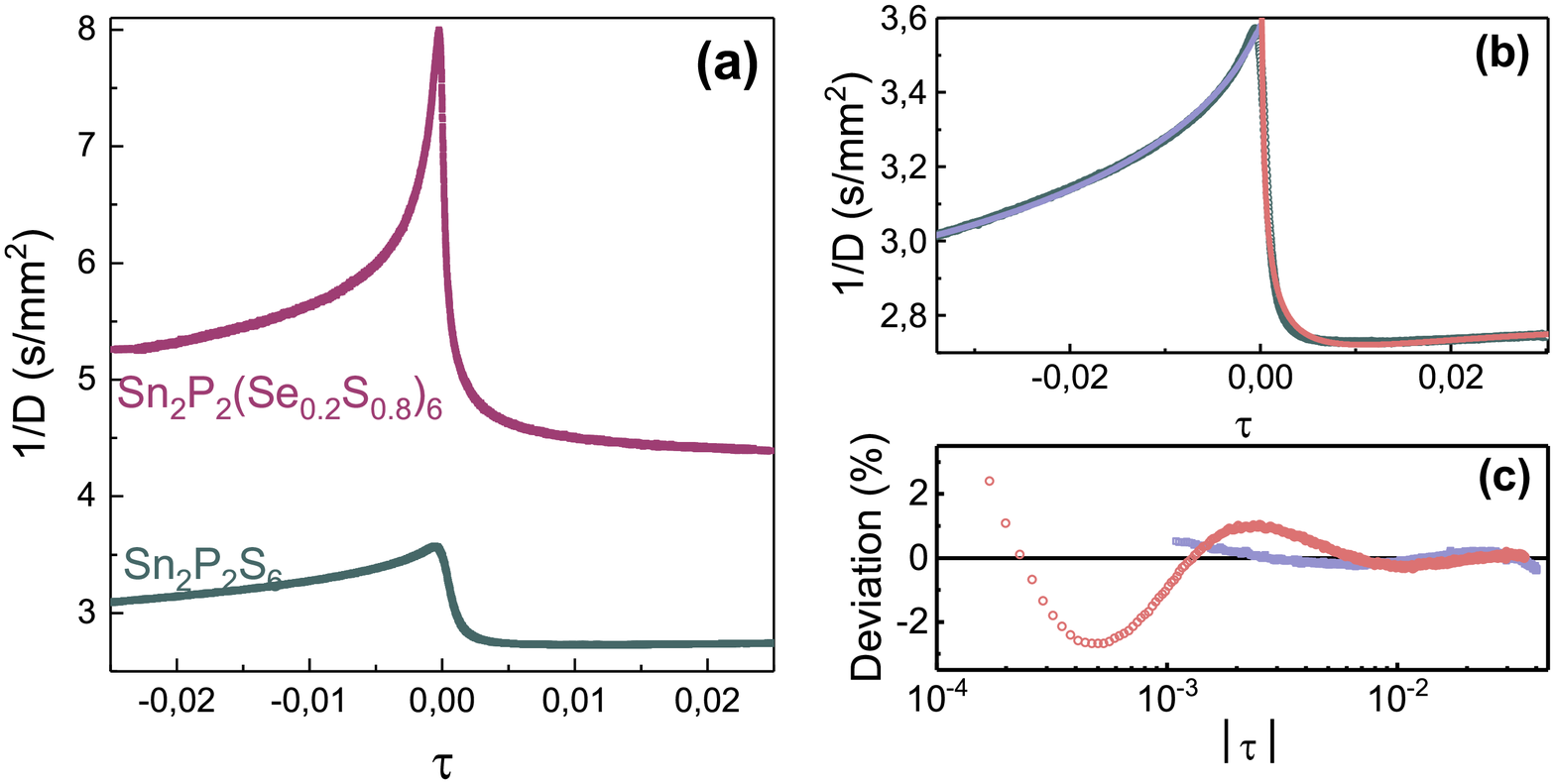}
	 \caption{(a) Dependence of the inverse of thermal diffusivity measured in the [100] crystallographic direction in the vicinity of continuous phase transition in \SPS and Sn$_2$P$_2$(Se$_{0.2}$S$_{0.8}$)$_6$ crystals;\cite{b2019_43,b2019_44} (b) results of the fittings of the inverse of thermal diffusivity for \SPS crystal, and (c) coresponding deviation curves for the fittings. The points correspond to experimental measurements, the continuous lines to the fittings to Eq.~\ref{eq4}. Blue color corresponds to the fitting below the critical temperature, red to the ones above it. \label{fig7}}
\end{figure*}

According to previous investigations of thermal diffusivity,\cite{b2019_43} which is proportional to the reciprocal of heat capacity, the critical behavior of \SPS can be fitted with a good quality in the paraelectric phase to a model which considers the influence of both fluctuations and defects. On the other hand, below $T_0$ the best fittings were found using a mean field Landau model. Such asymmetry is very strange, especially when it has been possible to describe the critical behavior of mixed crystals based on \SPS using a single model for both phases. For \SPSS solutions with increasing selenium concentration and approaching the LP ($x\approx 0.28$), the critical anomalies are nicely described by exponents and ratios of critical amplitudes that belong to the Lifshitz universality class [see Fig.~\ref{fig7}(a)].\cite{b2019_44} At tin by lead substitution in \PSPS mixed crystals, the critical behavior is also satisfactorily described above and below $T_0$ as a crossover from a clear non-mean field model at $y = 0.1$ to a mean field one at $y = 0.3$.\cite{b2019_20} For \SPSe crystals, at the paraelectric to incommensurate second order phase transition the critical anomaly above and below $T_i$ agrees with the predictions of the renormalization group theory for 3D-XY universality class.\cite{b2019_48} At simultaneous substitution of chemical elements in cation and anion sublattices in \PSPSS mixed crystals a TCLP has been found\cite{b2019_41} for $x = 0.28$ and $y = 0.05$ at $T_c = 259.12$~K. Here the tricritical Lifshitz universality class has been assigned by the obtained critical exponent $\alpha= 0.64$ which is equal to the theoretical predicted one.\cite{b2019_14}

Why a critical behavior of \SPS crystals can’t be described satisfactorily in both temperature sides of $T_0$? As was mentioned above, for pure \SPS crystals below $T_0$ the coexistence of antipolar (antiferroelectric-like) and ferroelectric states is possible. This is manifested by the observation of double hysteresis loops and usual ferroelectric loops.\cite{b2019_4} Considering the possibility of the coexistence of antipolar and ferroelectric phases below $T_0$, we reexamine previous experimental data.\cite{b2019_43} We have used for the fittings the well-known equation 
 \begin{equation}
\frac1D=B+C\tau+A^{\pm}|\tau|^{-\alpha}(1+E^{\pm}|\tau|^{0.5}),\label{eq4}
\end{equation}
\noindent where $\tau= (T - T_0)/T_0$ is the reduced temperature, and superscripts ``+'' and ``–'' stand for $T > T_0$ and $T < T_0$,
respectively, $\alpha$, $A^{\pm}$, $B$, $C$ and $E^{\pm}$ are adjustable parameters. Figure~\ref{fig7}(b) demonstrates the fittings together with the deviation plots [see Fig.~\ref{fig7}(c)] (difference between each experimental and fitted value, divided by the experimental value, in percentage).

The results of the fittings [see Fig.~\ref{fig7}(b),(c)] in \SPS have shown a XY-like behavior with the critical exponent $\alpha= - 0.0092 \pm 0.0008$ below $T_0$ and Ising like one with $\alpha= 0.1049 \pm 0.0066$ above $T_0$. Such value of the critical exponents, smaller than $\alpha_{\mathrm{ISING}}$ in the paraelectric phase and a little bigger than $\alpha_{\mathrm{XY}}$ below $T_0$, can be interpreted as a possible crossover in the critical behavior which earlier was predicted\cite{b2019_54} for systems with two competing order parameters near a bicritical point on the phase diagram. The bicritical point can be originated instead of a tetracritical point if there is a strong enough coupling of two order parameters.\cite{b2019_55}

In any case, good fittings of thermal diffusivity in \SPS crystal below $T_0$ with negative value of critical index demonstrate its “cups” shape that is a characteristic of the antiferroelectric-like ordering and coincides with observed double hysteresis loops in \SPS crystal below $T_0$.\cite{b2019_4}

\section{Conclusions}

Static and dynamic critical behavior of \SPS type ferroelectrics and \PSPSS mixed crystals are governed by multicritical points presence on their phase diagram. At hydrostatic compression of \SPS crystal or at tin by lead substitution in \PSPS mixed crystals the TCP can be reached, what is described by the BEG model for a system with three-well local potential for pseudospins fluctuations. At sulfur by selenium replacement in \SPSS solid solutions the LP is induced, what is explained by first and second neighbors short range interaction ratio changing in pseudospin ANNNI model. At simultaneous variation of chemical composition in cationic and anionic sublattices the lines of TCPs and LPs on $T - x - y$ phase diagram meet at the TCLP, and this higher order multicritical point can be described in a combined BEG-ANNNI model.\cite{b2019_37} Below the temperature of TCLP, which can be considered as the Lifshitz line end point, the “chaotic” state with coexisting ferroelectric and metastable paraelectric and modulated phases is possible. This expectation agrees with the concentration evolution of heat capacity\cite{b2019_48} and dielectric susceptibility\cite{b2019_49} temperature dependence in \PSPSe mixed crystals, which demonstrates a gradual lock-in transition smearing with growth of lead concentration. 

In addition to the frustration of polar fluctuations near the center of the Brillouin zone, in \SPS crystal the antipolar fluctuations also strongly develop in the paraelectric phase on cooling to the continuous phase transition temperature $T_0$. This is confirmed by the observation in neutron scattering experiments\cite{b2019_35} of a flat polar soft optical phonon branch and by the development of diffuse X-ray scattering along $q_y$ direction in the Brillouin zone of the paraelectric phase near temperature $T_0$.\cite{b2019_36} Such observation indicates closeness of this transition to a tetracritical point which appears due to the interaction between two order parameters, related to polar fluctuation near the Brillouin zone center and to antipolar fluctuations near its edge. At strong enough coupling of the mentioned order parameters, the tetracritical point can evolve to a bicritical point. The frequency dependence of the dielectric susceptibility temperature anomaly around $T_0$,\cite{b2019_52,b2019_53} together with observed aging effects\cite{b2019_56} and transformation of double hysteresis loops into usual ferroelectric-like loops\cite{b2019_4} confirm the possibility of simultaneous development of polar and antipolar fluctuations in the paraelectric phase on cooling to $T_0$, and the coexistence of antipolar and polar clusters in \SPS crystals below $T_0$. At this transition, the critical behavior according to the thermal diffusivity data\cite{b2019_43} can be described as a crossover between Ising and XY universality classes, what is expected near a bicritical point with coupled polar and antipolar order parameters.

\end{document}